%% file: main.tex
\documentclass[prl, unsortedaddress, superscriptaddress, twocolumn, twocolappendix,nobibnotes,reprint]{revtex4-2}
\usepackage{amsmath}
\usepackage{bm}
\usepackage{tensor}
\usepackage{booktabs}
\usepackage{graphicx}
\usepackage{xcolor}
\usepackage{amssymb}
\usepackage{hyperref}
\usepackage{aas}
\usepackage{orcidlink}

\hypersetup{
    colorlinks=true,   
    linkcolor=blue,    
    filecolor=blue,    
    citecolor=blue,    
    urlcolor=blue,     
    pdfborder={0 0 0}  
}

\usepackage[nameinlink,noabbrev]{cleveref}
\crefname{equation}{Eq.}{Eqs.}
\crefname{section}{Section}{Sections}
\crefname{figure}{Figure}{Figures}
\crefname{table}{Table}{Tables}
\crefname{appendix}{Appendix}{Appendices}
\Crefname{figure}{Figure}{Figures}
\Crefname{equation}{Equation}{Equations}
\Crefname{section}{Section}{Sections}
\Crefname{table}{Table}{Tables}

\renewcommand{\d}[0]{\mathrm{d}}

\newcommand{\kmsMpc}{\,{\rm km\,s^{-1}\,Mpc^{-1}}}

\newcommand{\uhm}{Department of Physics and Astronomy, University of Hawaii at M\=anoa, 2505 Correa Rd., Honolulu, HI, 96822 USA}
\newcommand{\asu}{School of Earth and Space Exploration, Arizona State University, Tempe, AZ 85287-6004 USA}
\newcommand{\bu}{Department of Physics, Boston University, 590 Commonwealth Avenue, Boston, MA 02215 USA}
\newcommand{\iac}{Instituto Avanzado de Cosmolog\'{\i}a A.~C., San Marcos 11 - Atenas 202. Magdalena Contreras. Ciudad de M\'{e}xico C.~P.~10720, M\'{e}xico}

\begin{document}

\author{S.~P.~Ahlen\orcidlink{0000-0001-6098-7247}}
\affiliation{\bu}

\author{A.~Aviles\orcidlink{0000-0001-5998-3986}}
\affiliation{Instituto de Ciencias F\'{\i}sicas, Universidad Nacional Aut\'onoma de M\'exico, Avenida Universidad s/n, Cuernavaca, Morelos, C.~P.~62210, M\'exico}
\affiliation{\iac}

\author{B.~Cartwright\orcidlink{0009-0003-6667-4729}}
\affiliation{\asu}

\author{K.~S.~Croker\orcidlink{0000-0002-6917-0214}}
\affiliation{\asu}
\affiliation{\uhm}

\author{W.~Elbers\orcidlink{0000-0002-2207-6108}}
\affiliation{Institute for Computational Cosmology, Department of Physics, Durham University, South Road, Durham DH1 3LE, United Kingdom}

\author{D.~Farrah\orcidlink{0000-0003-1748-2010}}
\affiliation{\uhm}
\affiliation{Institute for Astronomy, University of Hawaii,  2680 Woodlawn Drive, Honolulu, Hawaii, 96822 USA}

\author{N.~Fernandez\orcidlink{0000-0002-3573-339X}}
\affiliation{NHETC, Department of Physics and Astronomy, Rutgers University, Piscataway, New Jersey 08854 USA}

\author{G.~Niz\orcidlink{0000-0002-1544-8946}}
\affiliation{Departamento de F\'{\i}sica, DCI-Campus Le\'{o}n, Universidad de Guanajuato, Loma del Bosque 103, Le\'{o}n, Guanajuato C.~P.~37150, M\'{e}xico}
\affiliation{\iac}

\author{J.~W.~Rohlf\orcidlink{0000-0001-6423-9799}}
\affiliation{\bu}

\author{G.~Tarl\'e\orcidlink{0000-0003-1704-0781}}
\thanks{Contact author: gtarle@umich.edu}
\affiliation{Department of Physics, University of Michigan, 450 Church Street, Ann Arbor, Michigan 48109, USA}

\author{R.~A.~Windhorst\orcidlink{0000-0001-8156-6281}}
\affiliation{\asu}

\input{desi_collaboration_authors.tex}

\title{Positive Neutrino Masses with DESI DR2 via Matter Conversion to Dark Energy}

\begin{abstract}
  The Dark Energy Spectroscopic Instrument (DESI) is a massively parallel spectroscopic survey on the Mayall telescope at Kitt Peak, which has released measurements of baryon acoustic oscillations determined from over 14 million extragalactic targets. 
 We combine DESI Data Release 2 with CMB datasets to search for evidence of matter conversion to dark energy (DE), focusing on a scenario mediated by stellar collapse to cosmologically coupled black holes (CCBHs).
 In this physical model, which has the same number of free parameters as $\Lambda$CDM, DE production is determined by the cosmic star formation rate density (SFRD), allowing for distinct early- and late-time cosmologies.
 Using two SFRDs to bracket current observations, we find that the CCBH model: accurately recovers the cosmological expansion history, agrees with early-time baryon abundance measured by BBN, reduces tension with the local distance ladder, and relaxes constraints on the summed neutrino mass $\sum m_\nu$.
 For these SFRDs, we find a peaked positive $\sum m_\nu < 0.149\,\rm eV$ (95\% confidence) and $\sum m_\nu = 0.106^{+0.050}_{-0.069}\,\rm eV$, respectively, in good agreement with lower limits from neutrino oscillation experiments.
   A peak in $\sum m_\nu > 0$ results from late-time baryon consumption in the CCBH scenario and is expected to be a general feature of any model that converts sufficient matter to dark energy during and after reionization.
\end{abstract}

\maketitle

\emph{Introduction}---
The last three decades have seen a large and concerted effort to understand the observed late-time accelerating expansion of the Universe.
Early measurements of accelerating expansion were most easily understood in the context of a cosmologically constant energy density $\Lambda$.
When paired with cold dark matter (CDM), the $\Lambda$CDM model was successful in explaining a multitude of observations.
However, from a fundamental physics perspective, the measured value of $\Lambda$ made little sense~(e.g., Ref.~\cite{Weinberg1989}).
Now, recent observations have begun to raise tensions with the $\Lambda$CDM model ~(e.g., Ref.~\cite{divalentino21review}) and a consensus is emerging that $\Lambda$ should be understood as an approximation to a dynamical species termed dark energy (DE).

The first year results from the Dark Energy Spectroscopic Instrument (DESI DR1), using baryon acoustic oscillation (BAO) measurements from galaxies and quasars at $0.1 < z<2.1$ \cite{DESI2024.II.KP3,DESI2024.III.KP4} and auto- and cross-correlations of the Lyman-$\alpha$ forest and quasars at $2.1 < z < 4.2$ \cite{DESI2024.IV.KP6}, have found tentative evidence for DE dynamics. 
A follow-up full-shape (FS) analysis of the power spectrum supports the initial BAO results~\cite{DESI2024.V.KP5,DESI2024.VII.KP7B}.
The evidence for dynamic DE has strengthened in the just released results (DESI DR2) \cite{DESI2025.II,DESI2025.extde}, reaching $3.1 \sigma$ when combined with the cosmic microwave background (CMB) data from \emph{Planck} \cite{Planck2020cosmoparam} and the Atacama Cosmology Telescope (ACT) \cite{Madhavacheril_2024}.
The addition of supernovae (SNe) cosmology data gives evidence up to $4.2 \sigma$, depending on the dataset \cite{Scolnic_2022,Brout_2022,Riess_2022,rubin24union,desy52024}.

It is known that DE and neutrino mass can be degenerate~(e.g., Refs.~\cite{Hannestad2005, Dirian2017}), and tensions with DESI DR2 have also emerged with respect to the summed neutrino masses.
By convention, neutrino masses are labeled $m_1$, $m_2$ and $m_3$, with $m_1 < m_2$.
Neutrino oscillation measurements~\cite{navas24rev} show that at least two of the neutrinos have nonzero masses and that the squared mass differences $\Delta m_{ji}^2 := m_j^2 - m_i^2$, are given by $\Delta m_{21}^2 \sim 7.5 \times 10^{-5}\,\mathrm{eV}^2$ and $\left|\Delta m_{32}^2\right| \sim 2.5\times 10^{-3}\,\mathrm{eV}^2$. 
Thus, two of the masses, $m_1$ and $m_2$, are close to each other, while $m_3$ is either much larger than $m_1$ and $m_2$, referred to as normal ordering (NO), or much smaller than $m_1$ and $m_2$, referred to as inverted ordering (IO). 
It follows that the smallest summed mass allowed by normal (inverted) ordering is $0.05898 \pm 0.0003$~eV ($0.09982 \pm 0.0006$~eV).
DESI DR2, however, reports $\sum m_\nu < 0.0642~\mathrm{eV}$ at 95\% confidence~\cite{desi-neutrino-dr2}, in tension with these lower bounds.
When allowing an effective negative mass~\cite{planck14, alam21, craig24, ElbersFrenk2024,greenmeyer24}, this tension increases to $>3\sigma$.
Because neutrinos cannot acquire mass via the Higgs mechanism, the origin of neutrino mass is unknown.
Proposed mechanisms for producing neutrino mass can be distinguished by whether the mass eigenstates satisfy NO or IO. 
Measurement of the summed neutrino mass therefore gives crucial insight into new physics beyond the standard model.

Departures from $\Lambda$CDM are often parameterized using an effective DE equation of state $w_\mathrm{eff}(a)=w_0+w_a(1-a)$, where $a$ is the cosmological scale factor, and $w_0$ and $w_a$ are constants.
Although originally developed to study a single minimally coupled scalar field~\cite{ChevallierPolarski2001}, the $w_0w_a$ model has become a ``common ground,'' upon which specific physical models are interpreted as if they were a single species evolving cosmologically in isolation.
A salient feature of the best-fit DESI DR1 and DR2 $w_0w_a$CDM models is the presence of a ``phantom crossing,'' where the DE equation of state $w_\mathrm{eff} < -1$ becomes $w_\mathrm{eff} \geqslant -1$.
If taken literally, this feature of $w_0w_a$ is acausal but can be understood from the source-free conservation relation: $w_\mathrm{eff} = -1 - (a/3\rho)\,\d\rho/\d a$.
When physical densities in an expanding universe remain constant or dilute in time, then $w_\mathrm{eff} \geqslant -1$.
However, if the data demand that $\d\rho/\d a > 0$, then $w_\mathrm{eff}(a) < -1$.
In other words, $w_\mathrm{eff}$ appears phantom if stress energy is injected from other species.
Physical models for dynamic DE have been proposed in which baryonic or dark matter, or both, is converted to DE over cosmic time~(e.g., Refs.~\cite{Amendola2000, 2004ApJ...604....1F, PereiraJesus2009, CaiSu2010, PourtsidouSkordis2013, Yang:2018euj,Nunes:2022bhn}).
By construction, the energy injected into the DE species from matter gives increasing DE density over time, so the underlying physics remains causal. 
Furthermore, because these models exchange matter for DE, during cosmological parameter estimation they can relax bounds on the other non-relativistic species at late time: neutrinos. 

In this Letter, we examine one such model where conversion of baryons to DE occurs during the collapse of massive stars to regions of energized vacuum, i.e. nonsingular black holes (BHs)~\cite{czf2021,farrah2023b,croker24jcap,Cadoni2024a,Cadoni2024b}.
It has been shown that nonsingular BHs can become cosmologically coupled (CCBHs)~\cite{farja07,faraoni24}, contributing in aggregate as a DE species (as opposed to a matter species) with equation of state $w_\mathrm{phys} \gtrsim -1$~\cite{CrokerWeiner19, CrokerWeiner22}.
CCBH models that recover plausible expansion histories deplete baryonic density consistent with late-time censuses~(e.g., Refs.~\cite{Shull2012, Nicastro2018, Driver2021}) and are in harmony with both the DESI DR1 BAO measurements~\cite{croker24jcap} and with studies of supermassive black hole assembly in galaxies~\cite{farrah23a,lacy24,farrah25}.
Using data from DESI DR2 and CMB, we show that CCBH models fit as well as $\Lambda$CDM, while simultaneously giving physically reasonable values for $\sum m_\nu$, the baryon abundance today and at recombination, and an $H_0$ closer to local distance ladder measurements.
The End Matter details various technical points of our analysis.
Throughout, we assume spatial flatness and set $c := 1$ when convenient.

\emph{Theory}---
A number of exact general relativity (GR) solutions describing nonsingular BHs with $w_\mathrm{phys}=-1$ (energized vacuum) interiors have been found~(e.g., Refs.~\cite{BardeenNonsingular1968, Dymnikova92, mazur2004, Mbonye2005, Cattoen05, Lobo06, MazurMottola15, BeltracchiGondolo2019, KonoplyaPosada2019, BeltracchiPosada2024}).
The conversion of baryonic material at supranuclear densities into energized vacuum during stellar gravitational collapse is analogous to the time reversal of inflationary reheating~(e.g., Refs.~\cite{alb82, bassett06, allahverdi10, JeongKamada2023}).
It has been established from the Einstein-Hilbert action that ultracompact regions of energized vacuum can become cosmologically coupled~\cite{CrokerWeiner19, CrokerWeiner22}, impacting global dynamics in aggregate as a $w_\mathrm{phys} \gtrsim -1$ species.
Locally, each BH becomes tightly coupled to the embedding cosmology and grows in mass $m \propto a^k$, independently of accretion or merger.
The cosmological coupling strength $k := -3w_\mathrm{phys}$~\cite{czf2021} is determined in GR by the particular CCBH model solution adopted~\cite{crf2020}.
Models with $w_\mathrm{phys} = -1$ exhibit $m \propto a^3$ so that, modulo additional BH production, dilution in number density $\propto 1/a^3$ gives a cosmologically constant physical density.
Because BHs are formed via the collapse of massive stars, which have short lifetimes compared to the reciprocal expansion rate after star formation begins at $a_i \sim 1/20$, the DE density will grow, tracking the star formation rate~(c.f., Refs.~\cite{Afshordi2008,PrescodAfshordi2009}).

We model DE production via baryon consumption,
\begin{align}
  \rho_{\rm b} := \begin{cases}
    \displaystyle \frac{C\omega_{\rm b}^\mathrm{proj}}{a^3} & a < a_i \\
    \displaystyle \frac{C\omega_\mathrm{b}^\mathrm{proj}}{a^3} - \frac{\Xi}{a^3}\int_{a_i}^a \d a'\frac{\psi}{Ha'} & a \ge a_i
  \end{cases}\,,\label{eqn:rho_b_defn}
\end{align}
where $C$ is a dimensionful constant, $H$ is the Hubble rate, $\psi$ is the observed comoving star-formation rate density (SFRD), and
\begin{align}
  \omega_\mathrm{b}^\mathrm{proj} := C^{-1}\rho_\mathrm{b}(a_*)a_*^3.
\end{align}
Thus, $\omega_\mathrm{b}^\mathrm{proj}$ is the baryon density today that would be expected from measurements at recombination $a_*$, if there were no new physics at late times.
By definition, $\omega_\mathrm{b}$ is the baryon density at $a=1$, and so $\omega_\mathrm{b} < \omega_\mathrm{b}^\mathrm{proj}$ due to conversion of baryons into DE.
The single CCBH model parameter $\Xi$ encodes the initial amount of BH mass generated from baryon conversion, per unit of baryon mass.
During parameter estimation, $H_0$ (or equivalently the angular scale of the first CMB peak $100\theta_s$) becomes a derived parameter.
Thus, the CCBH model has the same number of free parameters as $\Lambda$CDM.
Conservation of stress-energy $\nabla^\mu T_{\mu\nu} = 0$ requires a source term from the baryons,
\begin{align}
  \frac{\d \rho_\mathrm{DE}}{\d a} = \frac{\Xi}{Ha^4}\psi\,. \label{eqn:ccbh_de_evolution}
\end{align}
Note that we have set $k:=3$ for the CCBH species, appropriate for the energized vacuum interiors of simple nonsingular BHs.
``Thawing'' occurs when $k < 3$, a scenario that we consider in the Discussion.

\emph{Methods}---
Our primary dataset is the DESI Data Release 2 (DR2).
DESI is a massively parallel 3.2-degree field spectroscopic survey on the Mayall telescope at Kitt Peak \cite{Snowmass2013.Levi, DESI2016b.Instr, DESI2022.KP1.Instr}.
It is capable of capturing nearly 5,000 simultaneous spectra from preselected targets in the Legacy Survey \cite{BASS.Zou.2017,LS.Overview.Dey.2019,TS.Pipeline.Myers.2023}, using robotic fiber positioners in an optically corrected focal plane \cite{FiberSystem.Poppett.2024, Corrector.Miller.2023,FocalPlane.Silber.2023}.
DESI's five-year observational campaign aims to collect more than 40 million galaxy and quasar spectra, along with spectroscopic classification and redshift measurements \cite{SurveyOps.Schlafly.2023,Spectro.Pipeline.Guy.2023,RedrockQSO.Brodzeller.2023}.
Publicly available catalogs~\footnote{\url{https://data.desi.lbl.gov/doc/releases/}} include the Early Data Release \cite{DESI2023b.KP1.EDR}, which followed a successful survey validation stage \cite{DESI2023a.KP1.SV}, and Data Release 1 (DR1)\cite{DESI2025.KP2.DR1}.

We augment these BAO data \cite{DESI2025.I, DESI2025.II} with \emph{Planck} PR4 CMB spectra likelihoods \textsc{lollipop} and \textsc{hillipop} (\texttt{L-H})~\cite{tristram24} for TT,TE,EE,lowE, and \textsc{commander} 2018 for lowTT.
Although slower than \textsc{camspec} \cite{Rosenberg2022}, foreground modeling in \texttt{L-H} allows inclusion of 100~GHz skymap data.
We will refer to this combination of likelihoods as ``Baseline''.
Because of possible microlensing systematics from CCBH along SNe lines of sight~\cite{2025MNRAS.536..946S, 2025MNRAS.537.3814S}, we do not include SNe datasets~\cite{desy52024, rubin24union, Scolnic_2022} in Baseline, but examine their impact in the Discussion.
We also exclude CMB lensing likelihoods from Baseline for technical reasons discussed in the End Matter.

To estimate the impact of uncertainties in the SFRD, we consider at $z > 4$ both the Trinca~\cite{trinca24} and Madau $\psi$~\cite{madau14,madaufragos2017}. 
These SFRDs are based on observational HST data and, in the case of Trinca $\psi$, JWST data and a matched semianalytic model.
For $z < 4$, the SFRD shape is well described by Madau $\psi$.
We normalize based on synchrotron emission in protostellar disks~\cite{hopkinsbeacom2006}, a measurement immune to obscuration effects.
These data, as a function of redshift, were converted to physical units (luminosities, space densities) using a fiducial $\Lambda$CDM with $\Omega_\mathrm{m} = 0.3$ and $H_0 = 70\,\kmsMpc$.
The impact of this fixed calibration is negligible for all cosmologies considered.
We adopt a single neutrino species with a degeneracy of 3 and a present-day temperature $T_\nu$ determined from the \emph{FIRAS} measurement of the CMB~\cite{fixsen09}.

We perform MCMC parameter estimation with \href{https://ui.adsabs.harvard.edu/abs/2019ascl.soft10019T}{\textsc{cobaya}}~\cite{TorradoLewis2021} using an augmented version of the Einstein-Boltzmann code \textsc{class} (further details are given in the End Matter).
The posterior distributions were computed using \href{https://getdist.readthedocs.io}{\textsc{getdist}}~\cite{Lewis2019}.
We evaluated the goodness of fit with $\chi^2$ from \textsc{bobyqa}~\cite{CartisFiala2018,CartisRobertsSheridan2018,Powell2009}, starting from the maximum {\it a posteriori} points (MAPs) of our MCMC chains.
Performance relative to $\Lambda$CDM is reported with $\Delta\chi^2_\mathrm{MAP} := -2\mathcal{L}_\mathrm{post}$ and reflects the impact of all priors.

\emph{Results}---
\begin{figure}
  \centering
  \includegraphics[width=\linewidth]{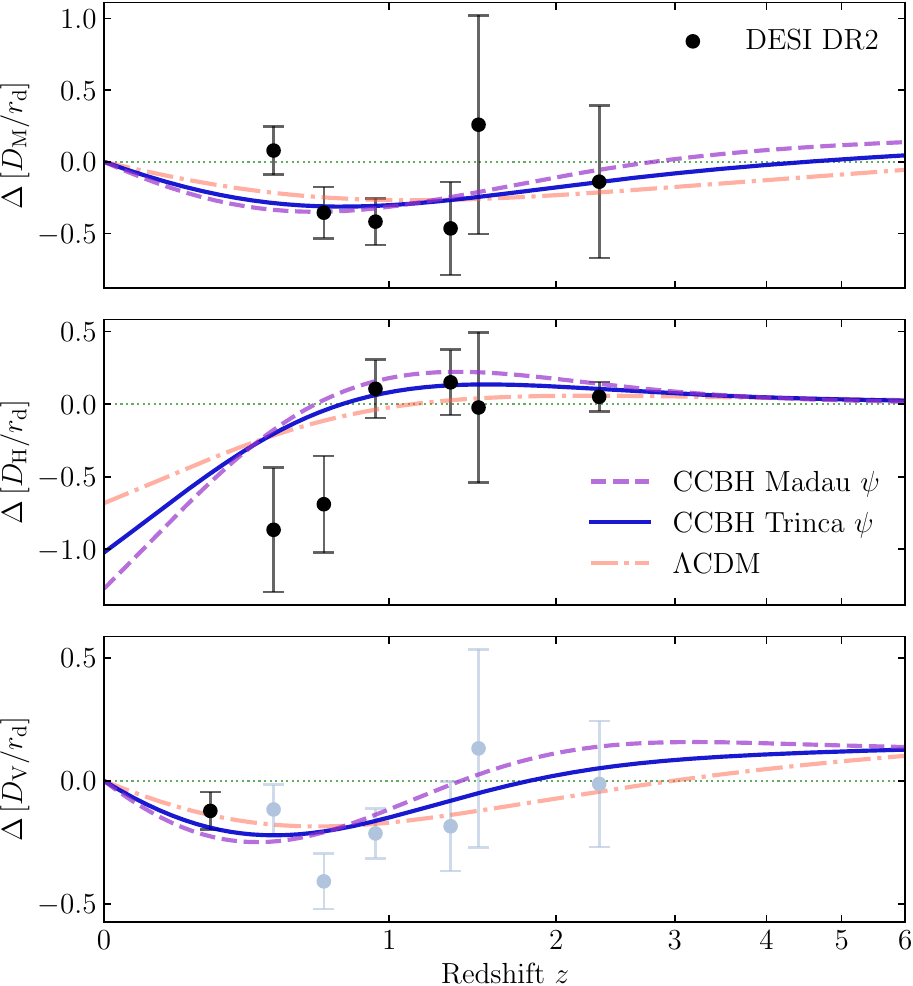}
  \caption{\label{fig:ahlenplot} Differences in BAO observables, relative to the {\it Planck} fiducial $\sum m_\nu := 0.06\,\mathrm{eV}$ $\Lambda$CDM cosmology.
    Displayed are transverse comoving distance $D_\mathrm{M}$, Hubble distance $D_\mathrm{H} \propto 1/H$, and volume-averaged distance $D_\mathrm{V}$ in units of the sound horizon at the end of baryon drag $r_\mathrm{d}$.
    DESI DR2 BAO measurements (black, 68\% confidence) are used to compute $D_\mathrm{V}$ values (light blue).
    We show CCBH Trinca $\psi$ (blue, solid line), Madau $\psi$ (purple, dashed line), and $\Lambda$CDM (orange, dash-dotted line) fit to DESI DR2 + CMB with floating $\sum m_\nu$.}
\end{figure}%
\begin{table}[b]
  \centering
  \def\arraystretch{1.2}
  \small   
  \begin{tabular}{lcccc}
   \toprule
   \midrule
    SFRD & $\Delta\chi^2_\mathrm{MAP}$  & $H_0$ [km/s/Mpc]& $\sum m_\nu$ [eV] & $\omega_\mathrm{b}/\omega_\mathrm{b}^\mathrm{proj}$ \\
   \midrule
   Trinca $\psi$ & $0.7$ & $69.37 \pm 0.36$ & $< 0.149$ (95\%) & $0.74^{+0.01}_{-0.03}$ \\
   Madau $\psi$ & $6.1$ & $70.03 \pm 0.40$ & $0.106^{+0.050}_{-0.069}$ & $0.50^{+0.03}_{-0.04}$ \\   
   \midrule
   \bottomrule
   \end{tabular}
  \caption{\label{tbl:results} Goodness of fit and posterior distributions (68\%) for expansion rate $H_0$, summed neutrino mass $\sum m_\nu$, and baryon survival fraction $\omega_\mathrm{b}/\omega_\mathrm{b}^\mathrm{proj}$ for bracketing star formation rate densities in our Baseline analysis.
    We compute $\Delta \chi^2_\mathrm{MAP}$ relative to $\Lambda$CDM with the same combination of datasets and likelihoods (see Methods).}
 \vspace{0.1em}
\end{table}%
\cref{fig:ahlenplot} shows the DESI DR2 BAO data, expressed as deviation of transverse comoving distance $D_\mathrm{M}$, Hubble distance $D_\mathrm{H}$, and volume-averaged distance $D_\mathrm{V} := (D_\mathrm{H} D_\mathrm{M}^2z)^{1/3}$ from fiducial \emph{Planck} TT+TE+EE+lowE 2018 $\Lambda$CDM values.
This already suggests dynamical DE, because the projected $\Lambda$CDM does not match when refined with BAO.
The goodness of fit to these data and CMB, with marginalized posteriors for $H_0$, $\sum m_\nu$, and baryon survival fraction $\omega_\mathrm{b}/\omega_\mathrm{b}^\mathrm{proj}$ is given in \cref{tbl:results} (priors in the End Matter).
Relative to Baseline $\Lambda$CDM, we find that CCBH Trinca $\psi$ is statistically indistinguishable, while Madau $\psi$ is disfavored at $\sim 2\sigma$ based on $\Delta\chi^2_\mathrm{MAP}$.
This is driven primarily by BAO with a small contribution from the high-$\ell$ CMB.

\begin{figure}
  \centering
  \includegraphics[width=\linewidth]{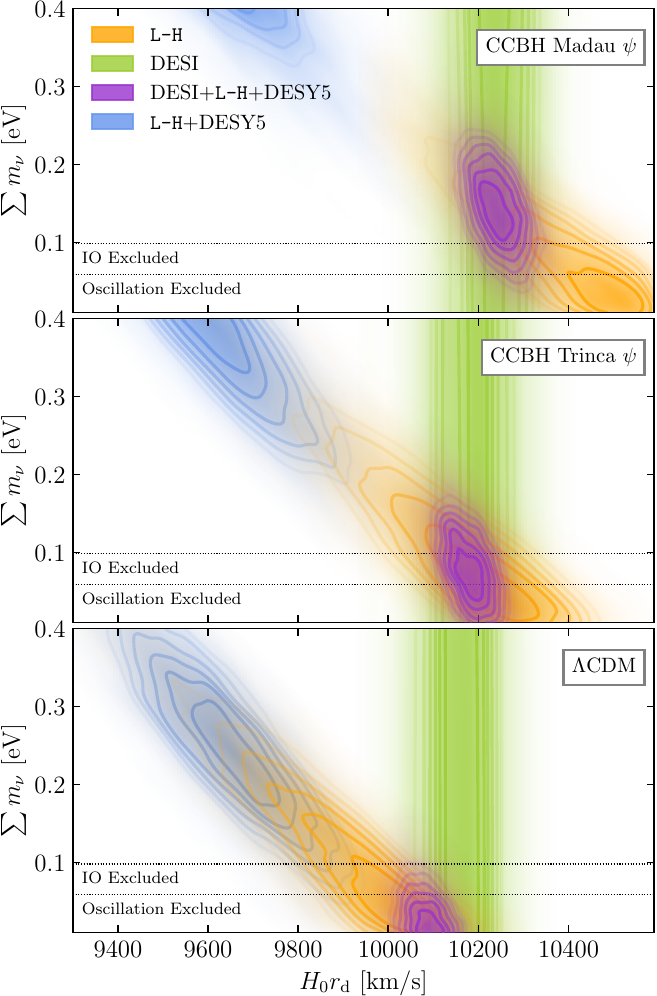}
  \caption{ \label{fig:H0rd_vs_Mtot} Consistency between DESI BAO (green) and CMB (orange) in summed neutrino mass $\sum m_\nu$ and the expansion rate today $H_0r_\mathrm{d}$.
    Probability density is shown via opacity and contours at $[0.2, 0.4, \dots, 1.4]\sigma$.
    The CMB centroid of $\Lambda$CDM (bottom) is offset from BAO.
    CCBH Trinca $\psi$ (middle) aligns posterior maxima, without degrading the posterior width, through more rapid transition to accelerated expansion via baryon conversion to DE.
    CCBH Madau $\psi$ (top) consumes twice as many baryons, coming into tension with BAO from the other side.
    In contrast to $\Lambda$CDM, inclusion of SNe (blue without BAO, purple with BAO) pulls $\sum m_\nu$ toward plausible values.
    DESY5 maximizes the effect for visualization.}
\end{figure}%
\begin{figure}
  \centering
  \includegraphics[width=\linewidth]{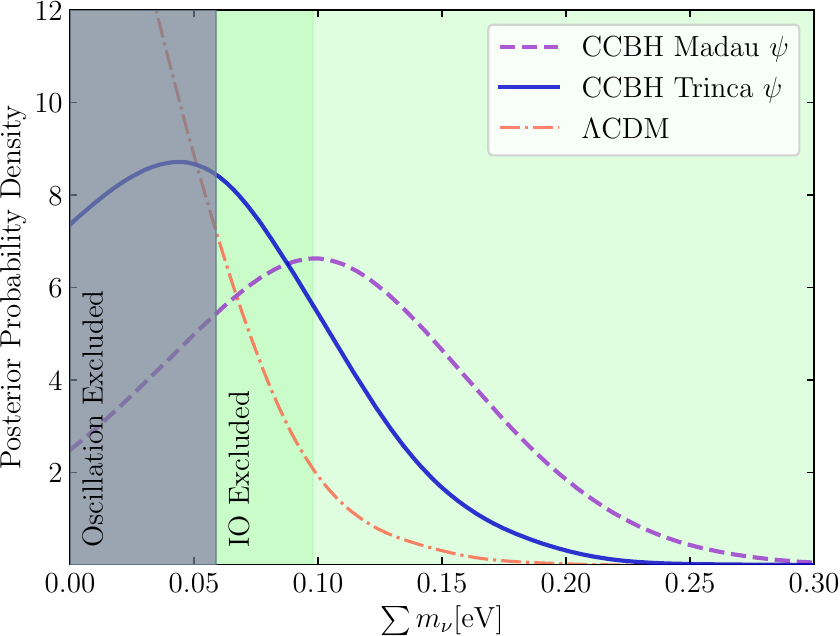}
  \caption{ \label{fig:nu-mass}
    Posterior probability densities for summed neutrino mass $\sum m_\nu$ under CCBH and $\Lambda$CDM (orange, dot-dashed line).
    Trinca $\psi$ (blue, solid line) and Madau $\psi$ (purple, dashed line) give peaks with $\sum m_\nu < 0.149$~(95\%) and $\sum m_\nu = 0.106^{+0.050}_{-0.069}$, respectively.
    We shade exclusions from neutrino oscillation experiments, indicating values consistent with the NO $m_1 < m_2 \ll m_3$ (dark green) and IO $m_3 \ll m_1 < m_2$ (light green).
    All fits use \emph{Planck} PR4 \texttt{L-H} CMB and DESI DR2 BAO.}
\end{figure}%

In \cref{fig:H0rd_vs_Mtot}, we show the consistency of CCBH with BAO and CMB in the $H_0r_\mathrm{d}$ -- $\sum m_\nu$ plane.
Unlike $\Lambda$CDM, CCBH scenarios can recover the same $H_0r_\mathrm{d}$ from CMB and BAO, independently, within a specific astrophysical model.
The CCBH framework allows for scenarios that can resolve the tension in $H_0r_\mathrm{d}$ between CMB and BAO measurements. 
As shown in \cref{fig:H0rd_vs_Mtot}, the CCBH Trinca $\psi$ model successfully aligns the posterior maxima from both datasets, resolving a tension that is present in the baseline $\Lambda$CDM analysis. 
The figure also demonstrates, however, that this alignment is sensitive to the star-formation history as the Madau $\psi$ model overcorrects this tension.
CCBH scenarios increase $H_0r_\mathrm{d}$ relative to Baseline $\Lambda$CDM because DE is added as matter is removed, speeding the transition to accelerated expansion.
We find Gaussian tensions with SH0ES' $H_0$ measurement~\cite{Riess_2022}, defined as the difference between mean values divided by their uncertainties summed in quadrature, i.e., $\left(H_{0}-H_{\mathrm{SH}0\mathrm{ES}}\right)/\sqrt{\sigma^{2}_{H_{0}} + \sigma^{2}_{\mathrm{SH}0\mathrm{ES}}}$, reduced from $4.2\sigma$ $\to$ $3.3\sigma$ (Trinca $\psi$) and $2.7\sigma$ (Madau $\psi$).

\cref{fig:nu-mass} shows posterior probability densities for the summed neutrino mass $\sum m_\nu$ under CCBH as compared to $\Lambda$CDM, a main result of this Letter. 
The central feature is that $\sum m_\nu$ for CCBH peaks at positive mass, about 0.05 eV (0.1 eV) for Trinca $\psi$ (Madau $\psi$), while $\Lambda$CDM peaks at negative mass. 
This occurs because conversion of matter to dark energy in CCBHs allows larger summed neutrino mass without changing the total matter density preferred by observations. 
The decrease in baryon density allows the model to remain consistent with the expansion history measured by DESI and CMB data while simultaneously resolving the tension on the summed neutrino mass seen in the $\Lambda$CDM analysis.
From \cref{eqn:ccbh_de_evolution}, DE production per unit baryon is more efficient at higher $z$.
Trinca $\psi$ features more abundant star formation at $z > 4$, and therefore requires less baryon consumption than Madau $\psi$. 
The higher value of the $\sum m_\nu$ posterior mass peak from Madau $\psi$ relative to Trinca $\psi$ is due to this $\sim 2\times$ higher consumption of baryons (see \cref{tbl:results}).
Trinca $\psi$ provides a novel explanation for the observational ``Missing Baryons Problem'' $\omega_\mathrm{b}/\omega_\mathrm{b}^\mathrm{proj} \sim 0.7$~\cite{Driver2021}, while the Madau scenario is at $\sim 2\sigma$ with respect to constraints from fast-radio bursts~\cite{Macquart2020}.
The actual SFRD is plausibly somewhere between Madau and Trinca $\psi$, which brackets the IO-excluded region.

\emph{Discussion}---
Similar to $\Lambda$CDM~(e.g., Ref.~\cite{DESI2025.II}, Fig.~10), the addition of SNe degrades $\Delta\chi^2_\mathrm{MAP}$ for both SFRDs (End Matter, \cref{tbl:sne_results}).
In the context of $w_0w_a$, this tension can be relieved by DE ``thawing,'' i.e., decaying in density at late times.
The CCBH model can accommodate similar internal dynamics if the individual BHs are not purely energized vacuum, so that $k < 3$.
Allowing $k$ to float in Baseline recovers $k = 2.74 \pm 0.14$, consistent with energized vacuum nonsingular BHs at $1.9\sigma$.
Including Pantheon+ SNe recovers $k = 2.70 \pm 0.10$ increasing the preference for thawing to $3\sigma$.
However, in both cases $H_0$ decreases, uncertainty in $\sum m_\nu$ increases, and a positive peak is lost.

In the CCBH scenario, neutrino masses can rise toward observational lower bounds because baryons at late time are converted into DE.
This allows neutrino masses to increase without decreasing the expansion rate.
In principle, similar physics should be present in models that convert CDM into DE during and after reionization.
We have focused on baryons because the CCBH scenario directly links the independently measured SFRD to the rate of baryon conversion to DE, whereas the internal dynamics of CDM remains relatively unconstrained.

\begin{table}
  \centering
  \def\arraystretch{1.2}
  \small   
  \begin{tabular}{lcccc}
   \toprule
   \midrule
   SFRD & $\Delta\chi^2_\mathrm{MAP}$ & $H_0$ [km/s\,Mpc] & $\sum m_\nu\,\mathrm{[eV]}$ & SNe \\
   \midrule
   Trinca $\psi$ & $5.14$ & $69.21^{+0.38}_{-0.34}$ & $0.075^{+0.022}_{-0.072}$ & Union3 \\ 
    & $6.28$ & $69.16 \pm 0.35$ & $0.076^{+0.028}_{-0.067}$ & Pantheon+ \\ 
    & $9.65$ & $68.97 \pm 0.36$ & $0.088^{+0.040}_{-0.066}$ & DESY5 \\ 
   \midrule
   Madau $\psi$ & $14.3$ & $69.80\pm0.40$ & $0.121^{+0.054}_{-0.065}$ & Union3 \\
    & $16.4$ & $69.71 \pm 0.39$ & $0.127^{+0.057}_{-0.063}$ & Pantheon+ \\ 
    & $21.8$ & $69.49 \pm 0.39$ & $0.144 \pm 0.06$ & DESY5 \\ 
   \midrule
   \bottomrule
   \end{tabular}
  \caption{\label{tbl:sne_results} Goodness of fit and posterior distributions (68\%) for $H_0$ and $\sum m_\nu$, for our Baseline analysis plus supernovae.  
    We compute $\Delta \chi^2_\mathrm{MAP}$ relative to $\Lambda$CDM with the same combination of datasets and likelihoods (see Methods).}
 \vspace{0.1em}
\end{table}%
\emph{Summary}---
DESI DR2 has provided compelling evidence for the evolution of DE.
The conversion of matter to DE allows the summed neutrino mass to increase while increasing $H_0$. 
Using two measured SFRDs as approximate bounds on the rate of star formation, we fit the CCBH model to DESI DR2 and \emph{Planck} PR4 data. 
With the same number of free parameters and goodness of fit as $\Lambda$CDM, the CCBH model recovers positive peaks in the summed neutrino mass consistent with the lower bounds from neutrino oscillation experiments. 
\par
\emph{Data availability}---
The data that support the findings of this Letter are openly available~\footnote{\url{https://github.com/CobayaSampler/bao_data/tree/master/desi_bao_dr2}}.
Data points in the figures are available in a Zenodo repository~\footnote{\url{https://doi.org/10.5281/zenodo.15734248}}.

\begin{acknowledgments}  
{\emph{Acknowledgments}---} We wish to thank the anonymous referee for carefully reading the manuscript and providing thoughtful suggestions that have improved the clarity of presentation.  
K.~C.~thanks Christopher Cain~(ASU) for guidance with computational resources.
W.~E.~acknowledges STFC Consolidated Grant No.~ST/X001075/1.
The work of NF is supported by DOE grant DE-SC0010008. 
G.~N.~and A.~A.~acknowledge the support of SECIHTI (Grant ``Ciencia Básica y de Frontera'' No. CBF2023-2024-162), DAIP-UG, and the Instituto Avanzado de Cosmologia. 
A.~A.~acknowledges support by PAPIIT IA101825 and PAPIIT IG102123. 
J.~W.~R.~acknowledges funding from US Department of Energy grant No.~DE-SC0016021.
G.~T.~acknowledges support through DoE Award No.~DE-SC009193.
R.~A.~W.~acknowledges support from NASA JWST Interdisciplinary Scientist Grants No.~NAG5-12460, No.~NNX14AN10G and No.~80NSSC18K0200 from GSFC. 

This work used Bridges-2 RM at the Pittsburgh Supercomputing Center through allocations PHY240332 and PHY250106 from the Advanced Cyberinfrastructure Coordination Ecosystem: Services \& Support (ACCESS)~\cite{boerner2023access} program, which is supported by U.S. National Science Foundation Grants No.~\#2138259, No.~\#2138286, No.~\#2138307, No.~\#2137603, and No.~\#2138296.

This material is based on work supported by the U.S. Department of Energy (DOE), Office of Science, Office of High-Energy Physics, under Contract No. DE–AC02–05CH11231, and by the National Energy Research Scientific Computing Center, a DOE Office of Science User Facility under the same contract. 
Additional support for DESI was provided by the U.S. National Science Foundation (NSF), Division of Astronomical Sciences under Contract No. AST-0950945 to the NSF’s National Optical-Infrared Astronomy Research Laboratory; the Science and Technology Facilities Council of the United Kingdom; the Gordon and Betty Moore Foundation; the Heising-Simons Foundation; the French Alternative Energies and Atomic Energy Commission (CEA); the National Council of Humanities, Science and Technology of Mexico (CONAHCYT); the Ministry of Science, Innovation and Universities of Spain (MICIU/AEI/10.13039/501100011033), and by the DESI Member Institutions \footnote{\url{https://www.desi.lbl.gov/collaborating-institutions}}. 
The authors are honored to be permitted to conduct scientific research on I'oligam Du'ag (Kitt Peak), a mountain with particular significance to the Tohono O’odham Nation.
\end{acknowledgments}
Any opinions, findings, conclusions, or recommendations expressed in this material are those of the author(s) and do not necessarily reflect the views of the U.S. National Science Foundation, the U.S. Department of Energy, or any of the listed funding agencies.

\bibliographystyle{apsrev4-2}
\bibliography{main_abbrev}{}
\appendix
\onecolumngrid
\section*{End Matter}
\twocolumngrid
In the absence of a preferred first-order model for the evolution of CCBH DE perturbations, the version of \textsc{class} we have used pins the CCBH fractional density perturbation to zero.
This is a reasonable approximation for the eventual spatial distribution of CCBHs, which are expected to disperse in aggregate at scales $\lesssim k_\mathrm{NL}$ \cite{crf2020}.
We note that any nontransient instability at first order would exclude a CCBH scenario.
Our study should be interpreted as applying to CCBH species with an effective sound speed squared $c_\mathrm{s}^2 \simeq 1/3$, as expected from a population of rapidly spinning CCBHs~\cite[][\S4.4]{crf2020}.

Our omission introduces a systematic error in the computation of the Newtonian gravitational potential $\psi$, because the baryon density perturbation is diminished by background baryon depletion without any initial compensating DE perturbation.
This can underestimate the total density perturbation at $z < 4$ by at most (Madau $\psi$) $\omega_\mathrm{b}/2\omega_\mathrm{bc} = 7.5\%$, conservatively.
One might expect that this diminished gravitational potential in the late-time universe contributes less lensed smoothing of the CMB, and so would shift $A_\mathrm{L} < 1$.
\begin{figure}[b]
  \centering
  \includegraphics[width=\linewidth]{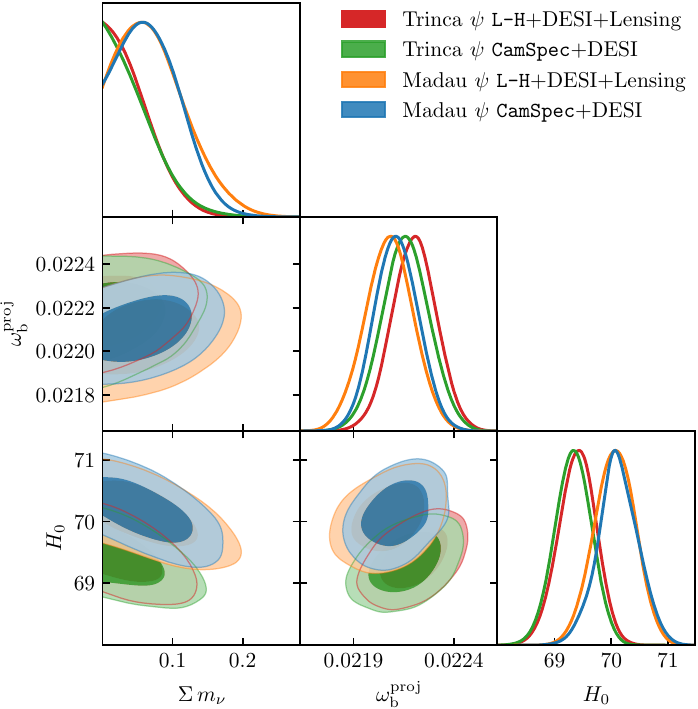}
  \caption{\label{fig:cmb_likelihoods} Impact of variations in CMB likelihood on summed neutrino mass $\sum m_\nu$, projected baryon density $\omega_\mathrm{b}^\mathrm{proj}$, and expansion rate today $H_0$.
    We consider CMB lensing (four-point constraint) and \textsc{camspec}, which handles skymap foregrounds differently than Baseline's \texttt{L-H}.
    Because of approximations within our Einstein-Boltzmann code, high $\ell$ multipoles will be increasingly inaccurate where CMB lensing has greatest constraining power.
    The effect is to reduce $\sum m_\nu$ by $\sim 0.04$~eV for both SFRDs.
    The impact of the \textsc{camspec} likelihood is nearly identical to \texttt{L-H} with CMB Lensing.}
\end{figure}%

We test whether more accurate tracking is required by floating $A_\mathrm{L}$ while fitting our Baseline to $\Lambda$CDM, Trinca $\psi$, and Madau $\psi$.
We find $A_\mathrm{L}$ equal to $1.068^{+0.049}_{-0.057}$ ($\Lambda$CDM), $1.076^{+0.055}_{-0.064}$ (Trinca), $1.101\pm0.066$ (Madau), all within $1-2\sigma$ of $A_\mathrm{L} = 1$.
Although consistent, movement in $A_\mathrm{L}$ is opposite to that expected.
This may result from late-time baryon consumption increasing the ratio $\omega_\mathrm{c}/\omega_\mathrm{b}^\mathrm{proj}$ in the overall fit, relative to $\Lambda$CDM.
This quantity roughly tracks potential well depth before recombination, because baryons have pressure but CDM does not.
We find that $\omega_\mathrm{c}/\omega_\mathrm{b}^\mathrm{proj}$ tracks $A_\mathrm{L}$.
So Madau has the deepest wells, implying more smoothing, and so the greatest $A_\mathrm{L}$.
A full-shape study, where the CCBH density perturbation is sourced from the baryon overdensity and gauge terms related to energy injection at background order, as well as a properly sinked baryon density contrast, will be presented in a second paper.

 In all analyses, we set $N_\mathrm{ur} := 0.00441$ as correct for \textsc{class} (c.f. $N_\mathrm{ur} = 0.00641$ as generated by \textsc{cobaya-cosmo-generator}), and adopted the following priors: $\omega_\mathrm{b}^\mathrm{proj} \in \mathcal{U}[0.005, 0.1]$; $\omega_\mathrm{c} \in \mathcal{U}[0.001, 0.99]$; $\tau_\mathrm{reio} \in \mathcal{U}[0.01, 0.8]$; $n_\mathrm{s} \in \mathcal{U}[0.8, 1.2]$; $\ln_{10} A_\mathrm{s} \in \mathcal{U}[1.61, 3.91]$; $m_\nu \in \mathcal{U}[0, 1.667]$~[eV]; $\omega_\Lambda \in \mathcal{U}[0.01, 0.99]$ ($\Lambda$CDM-only); $\Xi \in \mathcal{U}[0.87, 1.85]$ (CCBH Trinca $\psi$); and $\Xi \in \mathcal{U}[5.26, 6.23]$ (CCBH Madau $\psi$).
We ran four chains until a Gelman-Rubin $R-1 < 0.025$ was obtained and then burned $30\%$ of these chains to obtain posterior samples.
Priors for $\Xi$ and $\omega_\Lambda$ have been set to the same width to allow a meaningful comparison of $\Delta\chi^2_\mathrm{MAP}$ between models. 
Because $\Lambda$CDM is not a parameter limit of CCBH, and the two models have the same number of degrees of freedom, the quoted $\sigma$ from $\Delta\chi^2_\mathrm{MAP}$ is determined from the likelihood ratio $\exp(-\Delta\chi^2_\mathrm{MAP}/2)$.

Einstein-Boltzmann (EB) codes typically define the early-universe densities in terms of a present-day energy scale set by $H_0$.
In the CCBH scenario, $H_0$ is dynamically determined by stellar evolution physics, so it is not known \emph{a priori}.
We have adjusted a version of the EB code \textsc{class}~\cite{blas2011cosmic} to operate ``$h$-less,'' using the projected physical densities $\omega_i^\mathrm{proj}$ in place of the critical density fractions at the present epoch $\Omega_i$.
Within \textsc{class}, background quantities are computed on grids and then interpolated.
However, often within baryon-dependent code, an explicit scaling $\propto 1/a^3$ was implemented for speed, instead of doing table interpolation.
This becomes incorrect when baryons are depleted, so we have shifted most calculations to table interpolations. 
An exception is the second and third derivatives of the optical depth, which we retained for simplicity.
The extant calculations remain correct in the lead-up to, during, and after recombination.
However, interpolation during reionization will be slightly less accurate.
Similarly, the nonlinear codes for \textsc{halofit} and \textsc{hmcode} were not stripped of projection assumptions, so we have omitted lensing (four-point function constraint) from Baseline.

To gauge the robustness of our results, we investigated the sensitivity of both SFRDs to SNe, CMB lensing four-point constraint, and CMB TTTEEE likelihood.
The results for SNe are shown in \cref{tbl:sne_results}, where all SNe datasets degrade the quality of the fit.
This similarity to $\Lambda$CDM is expected because the CCBH scenarios mimic $\Lambda$CDM (e.g.~\cref{fig:ahlenplot}).
We investigated whether degradation was driven by $z < 0.1$ SNe, but found little sensitivity.

The impact of CMB likelihoods is displayed in \cref{fig:cmb_likelihoods}.
In general, we expect degradation of performance both because we explicitly omit nonlinear corrections, and perturbations at small scales are less accurate due to the instantaneous dispersal approximation.
For CMB, we consider two variations.
First, we replace \texttt{L-H} with PR4 NPIPE \textsc{camspec}~\cite{camspec21} for TTTEEE and \textsc{simall} for lowE. 
Second, we study the impact of \emph{Planck} PR4 lensing on \texttt{L-H}.
We find that the impact of \textsc{camspec} and CMB Lensing is nearly identical, shifting posterior $\sum m_\nu$ for Trinca $\psi \sim -0.05\,\mathrm{eV}$ and Madau $\psi \sim -0.04\,\mathrm{eV}$.
Increased $\sum m_\nu$ decreases potential wells during structure formation, which reduces the amount of smoothing due to gravitational lensing, implying lower $A_\mathrm{L}$.
It has been shown that \texttt{L-H} have a lower $A_\mathrm{L}$ compared to \textsc{camspec}~\cite[][Figure 18]{tristram24}, which suggests some of the offset between \cref{fig:nu-mass} and \cref{fig:cmb_likelihoods} could be due to $A_\mathrm{L}$.
The interplay between $\sum m_\nu$ and $A_\mathrm{L}$ will be investigated in future work through a full-shape analysis.

\end{document}

%% file: desi_collaboration_authors.tex

\author{J.~Aguilar}
\affiliation{Lawrence Berkeley National Laboratory, 1 Cyclotron Road, Berkeley, California 94720, USA}

\author{U.~Andrade\orcidlink{0000-0002-4118-8236}}
\affiliation{Leinweber Center for Theoretical Physics, University of Michigan, 450 Church Street, Ann Arbor, Michigan 48109-1040, USA}
\affiliation{Department of Physics, University of Michigan, 450 Church Street, Ann Arbor, MI 48109, USA}

\author{D.~Bianchi\orcidlink{0000-0001-9712-0006}}
\affiliation{Dipartimento di Fisica ``Aldo Pontremoli'', Universit\`a degli Studi di Milano, Via Celoria 16, I-20133 Milano, Italy}
\affiliation{INAF-Osservatorio Astronomico di Brera, Via Brera 28, 20122 Milano, Italy}

\author{D.~Brooks}
\affiliation{Department of Physics \& Astronomy, University College London, Gower Street, London, WC1E 6BT, United Kingdom}

\author{T.~Claybaugh}
\affiliation{Lawrence Berkeley National Laboratory, 1 Cyclotron Road, Berkeley, California 94720, USA}

\author{A.~de la Macorra\orcidlink{0000-0002-1769-1640}}
\affiliation{Instituto de F\'{\i}sica, Universidad Nacional Aut\'{o}noma de M\'{e}xico,  Circuito de la Investigaci\'{o}n Cient\'{\i}fica, Ciudad Universitaria, Cd. de M\'{e}xico  C.~P.~04510,  M\'{e}xico}

\author{A.~de~Mattia\orcidlink{0000-0003-0920-2947}}
\affiliation{IRFU, CEA, Universit\'{e} Paris-Saclay, F-91191 Gif-sur-Yvette, France}

\author{B.~Dey\orcidlink{0000-0002-5665-7912}}
\affiliation{Department of Astronomy \& Astrophysics, University of Toronto, Toronto, Ontario M5S 3H4, Canada}
\affiliation{Department of Physics \& Astronomy and Pittsburgh Particle Physics, Astrophysics, and Cosmology Center (PITT PACC), University of Pittsburgh, 3941 O'Hara Street, Pittsburgh, Pennsylvania 15260, USA}

\author{P.~Doel}
\affiliation{Department of Physics \& Astronomy, University College London, Gower Street, London, WC1E 6BT, United Kingdom}

\author{J.~E.~Forero-Romero\orcidlink{0000-0002-2890-3725}}
\affiliation{Departamento de F\'isica, Universidad de los Andes, Cra. 1 No. 18A-10, Edificio Ip, CP 111711, Bogot\'a, Colombia}
\affiliation{Observatorio Astron\'omico, Universidad de los Andes, Cra. 1 No. 18A-10, Edificio H, CP 111711 Bogot\'a, Colombia}

\author{E.~Gaztañaga}
\affiliation{Institut d'Estudis Espacials de Catalunya (IEEC), c/ Esteve Terradas 1, Edifici RDIT, Campus PMT-UPC, 08860 Castelldefels, Spain}
\affiliation{Institute of Cosmology and Gravitation, University of Portsmouth, Dennis Sciama Building, Portsmouth, PO1 3FX, United Kingdom}
\affiliation{Institute of Space Sciences, ICE-CSIC, Campus UAB, Carrer de Can Magrans s/n, 08913 Bellaterra, Barcelona, Spain}

\author{S.~Gontcho A Gontcho\orcidlink{0000-0003-3142-233X}}
\affiliation{Lawrence Berkeley National Laboratory, 1 Cyclotron Road, Berkeley, California 94720, USA}

\author{G.~Gutierrez}
\affiliation{Fermi National Accelerator Laboratory, PO Box 500, Batavia, Illinois 60510, USA}

\author{D.~Huterer\orcidlink{0000-0001-6558-0112}}
\affiliation{Department of Physics, University of Michigan, 450 Church Street, Ann Arbor, MI 48109, USA}

\author{M.~Ishak\orcidlink{0000-0002-6024-466X}}
\affiliation{Department of Physics, The University of Texas at Dallas, 800 W. Campbell Road, Richardson, Texas 75080, USA}

\author{R.~Kehoe}
\affiliation{Department of Physics, Southern Methodist University, 3215 Daniel Avenue, Dallas, Texas 75275, USA}

\author{D.~Kirkby\orcidlink{0000-0002-8828-5463}}
\affiliation{Department of Physics and Astronomy, University of California, Irvine, 92697, USA}

\author{A.~Kremin\orcidlink{0000-0001-6356-7424}}
\affiliation{Lawrence Berkeley National Laboratory, 1 Cyclotron Road, Berkeley, California 94720, USA}

\author{O.~Lahav}
\affiliation{Department of Physics \& Astronomy, University College London, Gower Street, London, WC1E 6BT, United Kingdom}

\author{C.~Lamman\orcidlink{0000-0002-6731-9329}}
\affiliation{Center for Astrophysics $|$ Harvard \& Smithsonian, 60 Garden Street, Cambridge, Massachusetts 02138, USA}

\author{M.~Landriau\orcidlink{0000-0003-1838-8528}}
\affiliation{Lawrence Berkeley National Laboratory, 1 Cyclotron Road, Berkeley, California 94720, USA}

\author{L.~Le~Guillou\orcidlink{0000-0001-7178-8868}}
\affiliation{Sorbonne Universit\'{e}, CNRS/IN2P3, Laboratoire de Physique Nucl\'{e}aire et de Hautes Energies (LPNHE), FR-75005 Paris, France}

\author{M.~E.~Levi\orcidlink{0000-0003-1887-1018}}
\affiliation{Lawrence Berkeley National Laboratory, 1 Cyclotron Road, Berkeley, California 94720, USA}

\author{M.~Manera\orcidlink{0000-0003-4962-8934}}
\affiliation{Departament de F\'{i}sica, Serra H\'{u}nter, Universitat Aut\`{o}noma de Barcelona, 08193 Bellaterra (Barcelona), Spain}
\affiliation{Institut de F\'{i}sica d’Altes Energies (IFAE), The Barcelona Institute of Science and Technology, Edifici Cn, Campus UAB, 08193, Bellaterra (Barcelona), Spain}

\author{R.~Miquel}
\affiliation{Instituci\'{o} Catalana de Recerca i Estudis Avan\c{c}ats, Passeig de Llu\'{\i}s Companys, 23, 08010 Barcelona, Spain}
\affiliation{Institut de F\'{i}sica d’Altes Energies (IFAE), The Barcelona Institute of Science and Technology, Edifici Cn, Campus UAB, 08193, Bellaterra (Barcelona), Spain}

\author{J.~Moustakas\orcidlink{0000-0002-2733-4559}}
\affiliation{Department of Physics and Astronomy, Siena College, 515 Loudon Road, Loudonville, New York 12211, USA}

\author{I.~P\'erez-R\`afols\orcidlink{0000-0001-6979-0125}}
\affiliation{Departament de F\'isica, EEBE, Universitat Polit\`ecnica de Catalunya, c/Eduard Maristany 10, 08930 Barcelona, Spain}

\author{F.~Prada\orcidlink{0000-0001-7145-8674}}
\affiliation{Instituto de Astrof\'{i}sica de Andaluc\'{i}a (CSIC), Glorieta de la Astronom\'{i}a, s/n, E-18008 Granada, Spain}

\author{G.~Rossi}
\affiliation{Department of Physics and Astronomy, Sejong University, 209 Neungdong-ro, Gwangjin-gu, Seoul 05006, Republic of Korea}

\author{E.~Sanchez\orcidlink{0000-0002-9646-8198}}
\affiliation{CIEMAT, Avenida Complutense 40, E-28040 Madrid, Spain}

\author{M.~Schubnell}
\affiliation{Department of Physics, University of Michigan, 450 Church Street, Ann Arbor, MI 48109, USA}

\author{H.~Seo\orcidlink{0000-0002-6588-3508}}
\affiliation{Department of Physics \& Astronomy, Ohio University, 139 University Terrace, Athens, Ohio 45701, USA}

\author{J.~Silber\orcidlink{0000-0002-3461-0320}}
\affiliation{Lawrence Berkeley National Laboratory, 1 Cyclotron Road, Berkeley, California 94720, USA}

\author{D.~Sprayberry}
\affiliation{NSF NOIRLab, 950 North Cherry Avenue, Tucson, Arizona 85719, USA}

\author{M.~Walther\orcidlink{0000-0002-1748-3745}}
\affiliation{Excellence Cluster ORIGINS, Boltzmannstrasse 2, D-85748 Garching, Germany}
\affiliation{University Observatory, Faculty of Physics, Ludwig-Maximilians-Universit\"{a}t, Scheinerstrasse 1, 81677 M\"{u}nchen, Germany}

\author{B.~A.~Weaver}
\affiliation{NSF NOIRLab, 950 N. Cherry Ave., Tucson, AZ 85719, USA}

\author{R.~H.~Wechsler\orcidlink{0000-0003-2229-011X}}
\affiliation{Kavli Institute for Particle Astrophysics and Cosmology, Stanford University, Menlo Park, California 94305, USA}
\affiliation{Physics Department, Stanford University, Stanford, California 93405, USA}
\affiliation{SLAC National Accelerator Laboratory, 2575 Sand Hill Road, Menlo Park, California 94025, USA}

\author{H.~Zou\orcidlink{0000-0002-6684-3997}}
\affiliation{National Astronomical Observatories, Chinese Academy of Sciences, A20 Datun Road, Chaoyang District, Beijing, 100101, Peoples
Republic of China}

\collaboration{DESI Collaboration}